\documentstyle[aps,aippreprint,floats,epsf]{revtex}
\epsfverbosetrue
\begin{document}
\pagestyle{empty}
\rightline{hep-ph/9504416}
\bigskip
\title{Gauge Symmetries and Vector-Boson Self 
Interactions$^{\dag}$}\footnotetext{Presented
at the International Symposium on Vector-Boson
Self Interactions, UCLA, February 1--3, 1995.}

\author{S.~Willenbrock}

\address{Department of Physics\\University of Illinois\\1110 West Green Street\\
Urbana, IL   61801}

\maketitle

\begin{abstract}
I review why we believe the electromagnetic, strong, and weak 
interactions are gauge theories, and what this implies for the self 
interactions of the gauge bosons.  The modern point of view regarding 
non-renormalizable effective field theories is emphasized.
\end{abstract}

\def\PR #1 #2 #3 {Phys.~Rev.~{\bf #1}, #2 (#3)} 
\def\PRL #1 #2 #3 {Phys.~Rev.~Lett.~{\bf #1}, #2 (#3)}
\def\PRD #1 #2 #3 {Phys.~Rev.~D~{\bf #1}, #2 (#3)}
\def\PLB #1 #2 #3 {Phys.~Lett.~{\bf #1}, #2 (#3)}
\def\NPB #1 #2 #3 {Nucl.~Phys.~{\bf #1}, #2 (#3)}
\def\RMP #1 #2 #3 {Rev.~Mod.~Phys.~{\bf #1}, #2 (#3)}
\def\ZPC #1 #2 #3 {Z.~Phys.~C~{\bf #1}, #2 (#3)}

\section*{Introduction}

The discovery of the $W$ and $Z$ bosons at the CERN 
$Sp\bar pS$ in 1983 \cite{WZ} began the era of the weak vector boson. 
It opened with a bang, earning a Nobel prize for the discovery of the 
particles and the development of the machine that made it possible 
\cite{NOBEL}. 
That era is now in its maturity, with precision studies of $Z$ bosons 
taking place at the CERN LEP and SLAC SLC $e^+e^-$ colliders, and studies of 
both $Z$ and $W$ bosons taking place at the Fermilab Tevatron $p\bar p$ 
collider.

The era of weak boson pair production began more quietly about two years 
ago with the first $WZ$ event at the Tevatron, shown in 
Fig.~1.  We will hear at this meeting of the first direct evidence for the 
$WWZ$ interaction from the CDF Collaboration \cite{FUESS,CDF}. Soon we will 
see the production of 
$W^+W^-$ pairs at the CERN LEP II $e^+e^-$ collider \cite{BUS}, 
and large numbers of weak boson pairs will be provided by the 
CERN LHC \cite{WOM}. Future 
$e^+e^-$ colliders will further contribute to the study of $W^+W^-$ 
and $ZZ$ pairs at high energy \cite{BAR}.

Given the present situation, this is an appropriate time to ask two
questions: 

\begin{itemize}

\item What have we learned from the weak-boson era?

\item What can we learn from the era of weak-boson pair production?

\end{itemize}

The language for this discussion will be quantum field theory.  As far as 
we know, quantum field theory is the only possible way to wed 
quantum mechanics and special 
relativity.\footnote{There is also string theory, but at low 
energies this reduces to quantum field theory.}  More precisely, it is the
only formalism capable of simultaneously implementing the constraints of
Lorentz invariance, unitarity, analyticity, and cluster decomposition 
\cite{PHYSICA}.\footnote{Cluster decomposition is the requirement that 
scattering amplitudes factorize when two particles are separated by a 
large spacelike distance.}
Due to the well-known ultraviolet divergences of quantum 
field theory, it is unlikely that it is a valid description of nature to 
arbitrarily high energies.  
Thus we believe that at the energies currently available to us, 
nature must be described by an ``effective'' quantum field 
theory, even though we do not believe that quantum field theory is truly 
fundamental \cite{PHYSICA,EFT}.

\begin{figure}
\begin{center}
\epsfxsize=0.9\textwidth
\leavevmode
\epsfbox[108 143 484 539]{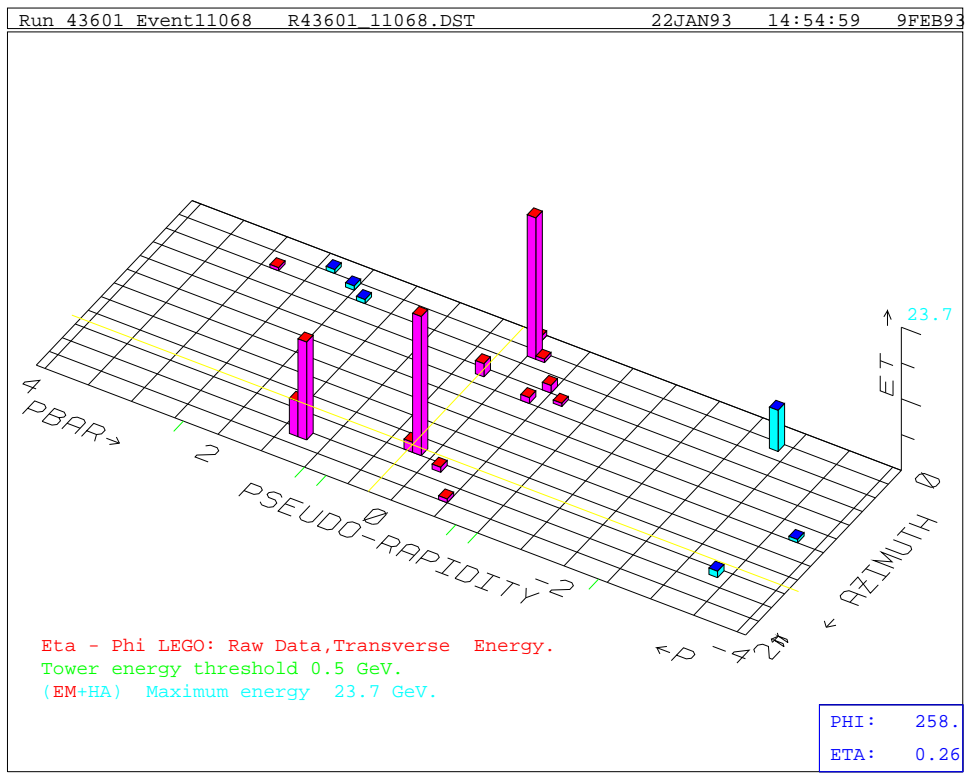}
%%\epsfbox[161 143 543 360]{wz_lego.ps}
\end{center}
\caption{The first $WZ$ event at the Fermilab Tevatron, with 
leptonic decay of both weak bosons. The two tallest towers are the $e^+e^-$
decay products of the $Z$ boson, and the third tallest tower is an $e^+$ 
from $W^+$ decay.}
\end{figure}

In this talk I present a discussion of vector-boson self interactions
from a modern point of view. The presentation is 
ahistorical, although I occasionally make remarks pertinent
to the historical development of the theory.  In particular, the modern 
point of view regarding non-renormalizable effective field theories figures 
prominently in the discussion.  My goal is to present a theoretical 
overview of the subject, and to point to subsequent speakers who will 
develop various subtopics in more detail.  In keeping with this style, I will 
often leave the citation of the literature to these 
speakers.\footnote{Some of the observations made in this talk are also made
in Ref.~\cite{EIN}.}

Although the subject of this talk is mostly of interest for the weak 
interaction, it is instructive to also consider the electromagnetic and 
strong interactions. 
The order of presentation is as follows:
\begin{itemize}

\item Quantum Electrodynamics

\item Quantum Chromodynamics

\item Weak Interaction:

      \begin{description}
      
      \item[$\circ$] Higgs model 

      \item[$\circ$] No-Higgs model 

      \end{description}

\end{itemize}
In a final section I reflect upon what we have learned from our 
deliberations. 

\section*{Quantum Electrodynamics}

\indent\indent Let us begin by building the theory of quantum 
electrodynamics from two experimental facts:
\begin{enumerate}

\item The photon is massless.\footnote{The experimental upper bound on the 
photon mass is 3$\times 10^{-27}$ eV.  For the sake of argument, let us 
regard the photon as being exactly massless.}

\item The photon has spin one.

\end{enumerate}
From these experimental facts, the challenge is to construct a 
consistent quantum field theory of photons and electrically-charged fermions. 
The simplest field which contains spin one is the vector 
field,\footnote{Tensor fields also contain spin one, but do not reproduce 
Maxwell's equations in the classical limit \cite{W5}.} so we 
begin by associating the photon with a field $A^{\mu}(x)$.  In so doing, 
we immediately encounter two difficulties:
\begin{enumerate}

\item The photon has only two degrees of freedom, corresponding to 
helicity $\pm 1$, while the vector field $A^{\mu}$ has four degrees of 
freedom.

\item The temporal component of the vector field has negative energy. 

\end{enumerate}
To see the latter point, consider the following Lagrangian for a vector 
field,
\begin{eqnarray}
{\cal L} & = & -\frac{1}{2}\partial^\mu A^\nu \partial_\mu A_\nu \\
         & = & -\frac{1}{2}\left[\left(\frac{\partial A^0}{\partial t}\right)^2
	       -\left(\frac{\partial {\bf A}}{\partial t}\right)^2
		+\cdots\right]		\nonumber
\end{eqnarray}
which shows that in order for the spatial components of the vector field 
to have positive energy, the temporal component must have negative 
energy. 

The resolution of these difficulties is well known.  To eliminate the 
negative-energy component, we add an additional term to the Lagrangian which 
cancels the offending term above,
\begin{eqnarray}
{\cal L} & = & -\frac{1}{2}(\partial^\mu A^\nu \partial_\mu A_\nu - 
\partial^\mu A^\nu \partial_\nu A_\mu) \\
         & = & -\frac{1}{2}\left[\left(\frac{\partial A^0}{\partial t}\right)^2
	 + \cdots - \left(\frac{\partial A^0}{\partial t}\right)^2 
	 + \cdots\right] \nonumber \\
         & = & -\frac{1}{4}F^{\mu\nu}F_{\mu\nu} \nonumber
\end{eqnarray}
where the last line casts the Lagrangian in the familiar form in terms of 
the electromagnetic field-strength tensor
\begin{equation}
F^{\mu\nu}= \partial^\mu A^\nu - \partial^\nu A^\mu \;.
\end{equation}
The field $A^0$ has been eliminated as a dynamical degree of freedom.
We now notice that this Lagrangian is invariant under the transformation
\begin{equation}
A^\mu \to A^\mu - \partial^\mu\lambda 
\label{GI}
\end{equation}
which allows us to eliminate another degree of freedom from the theory, 
bringing us down to the desired two degrees of freedom \cite{BD}. 

We recognize Eq.~\ref{GI} as the familiar gauge invariance of QED.  What 
the above argument shows in a heuristic way, and has been proven 
rigorously 
\cite{W5,Z}, is that gauge invariance is {\em mandatory}; it can be derived 
from 
the assumption of a massless spin one particle.\footnote{We now
regard gauge invariance as fundamental, and use it to {\em explain\/} the 
masslessness of the photon, the reverse of the above logic.  This  
point of view is largely a consequence of our realization that the strong 
and weak interactions are also gauge theories.} Gauge invariance is 
necessary to reconcile Lorentz invariance (the four-vector field 
$A^\mu$) and unitarity (two degrees of freedom).\footnote{An alternative 
point of view is that $A^\mu$ is not a four vector, because under Lorentz 
transformations it undergoes a gauge transformation as well. Again, gauge 
invariance is mandatory to ensure Lorentz invariance \cite{W5,VELTMAN}.} 

The necessity of gauge invariance in the formulation of QED implies that 
photon self interactions of the form 
\begin{equation}
{\cal L}_{int} = c_1 A^\mu A_\mu A^\nu A_\nu 
+ c_2 \partial^\mu A^\nu A_\mu A_\nu
\label{SELF}
\end{equation}
are strictly forbidden. Such terms are not gauge invariant, and their 
presence would destroy the consistency of the theory. 

This does not mean that there cannot be photon self interactions, however.
Let's write down the most general Lagrangian for the interaction of 
photons and fermions allowed by Lorentz invariance and gauge invariance:
\begin{eqnarray}
{\cal L} & = & -\frac{1}{4}F^{\mu\nu}F_{\mu\nu} +i\bar\psi \not\!\! D \psi - 
m\bar\psi\psi \label{QED} \\
         & + & \frac{c_1}{M^2}m\bar\psi\sigma^{\mu\nu}\psi F_{\mu\nu} + 
         \frac{c_2}{M^2}\bar\psi\gamma^\mu\psi\bar\psi\gamma_\mu\psi
	 \nonumber \\
         & + & \frac{c_3}{M^4}(F^{\mu\nu}F_{\mu\nu})^2 + \cdots \nonumber
\end{eqnarray}
where the terms are arranged in increasing powers of 
dimension,\footnote{QED possesses a global chiral 
symmetry, $\psi \to \exp[i\theta\gamma_5]\psi$, in the limit $m\to 0$, 
so we expect the coefficient of the term
$\bar\psi\sigma^{\mu\nu}\psi F_{\mu\nu}$, which violates this symmetry, to 
contain an explicit power of the fermion mass.}
and $M$ 
is a mass scale introduced to make the constants $c_i$ dimensionless. 
The first line above is the familiar Lagrangian of QED, and it describes 
the interaction of photons and fermions with remarkable success.  The 
(infinite number of) additional terms are unnecessary; there is no 
experimental observation which requires any of them. In the past, such 
terms would have been dismissed on the grounds that 
they are non-renormalizable; they have coefficients with inverse 
powers of mass, the hallmark of non-renormalizable interactions.  
However, we no longer regard 
renormalizability as a fundamental requirement of a field theory, since 
we do not demand that a given field theory (or even field theory itself) 
be valid to arbitrarily high 
energy.  Instead, we recognize that these additional terms are suppressed 
by inverse powers of $M$, which we regard as the energy scale at which 
ordinary QED ceases to be a valid description of the interaction of 
photons and fermions.  The presence of such terms would be revealed to us 
by performing experiments at sufficiently high energy or with sufficient 
accuracy.  The success of QED implies that $M$ is a very large mass, at 
least 1 TeV.  The renormalizability of ordinary QED ensures that these terms 
are not needed to cancel divergences, to all orders in perturbation theory, so 
the scale $M$ can be arbitrarily large. However, the renormalizability of
QED is just a consequence of the fact that $M$ is much larger than the 
currently accessible energy and accuracy.

The last term in Eq.~\ref{QED} represents a gauge-invariant 
four-photon interaction.\footnote{The term 
$F^\mu_\nu F^\nu_\rho F^\rho_\mu$ vanishes since $F^{\mu\nu}$ is 
antisymmetric.}  The observation of such an interaction would be 
evidence for new physics beyond QED, but would be consistent 
with what we already know about QED.

\section*{Quantum Chromodynamics}

\indent\indent Let us now approach QCD in a manner analogous to our 
approach to QED.  We again begin with a list of 
``experimental facts'':

\begin{enumerate}

\item The gluon is massless.\footnote{Since gluons (and quarks) are confined,
their masses cannot be measured directly.  There is no evidence for a 
bare gluon mass, so let us assume it is exactly massless. Gluons behave as 
if they have a dynamically-generated mass of order 300 MeV, in the same sense 
that quarks have a dynamically-generated ``constituent'' mass of the same 
order; this should not be confused with the bare mass.}  

\item The gluon has spin one.\footnote{As evidenced, for example, by the 
angular distribution of three-jet events in $e^+e^-$ collisions \cite{3JET}.}

\item The gluon interacts with itself.\footnote{This is necessary to explain
confinement, asymptotic freedom, and other phenomena.}

\end{enumerate}
Of course, these facts cannot be gleaned directly from experiment, 
which is the reason it took so many years to realize that QCD is the 
theory of the strong interaction. Let's construct a consistent theory 
which incorporates the above facts.

As with QED, we attempt to construct a theory based on the vector field 
$G^\mu(x)$.  We encounter the same difficulties as in QED (too many 
degrees of freedom, one of which has negative energy), 
with the same resolution (gauge invariance).  However, we argued that 
gauge invariance forbids vector-field self interactions, such as those in 
Eq.~\ref{SELF}, so we run into a new problem: how do we allow the 
gluon to interact with itself and not spoil gauge invariance?

The resolution of this problem is also well known.  Instead of a single 
gluon, we introduce a multiplet of gluons, eight to be exact.  We expand 
the gauge transformation of QED, Eq.~\ref{GI}, to include a rotation of 
the eight gluons into each other under the group SU(3).  The result is the 
familiar Yang-Mills theory of QCD, with eight self-interacting gluons.  The 
essential point is that, as in QED, gauge invariance is {\em mandatory\/} 
for the consistency of the theory.\footnote{To the best of 
my knowledge, it has never been rigorously shown that Yang-Mills gauge 
theory is the unique theory of massless, interacting, spin-one particles, 
based on vector fields. The necessity of gauge symmetry is suggested by the 
Weinberg-Witten theorem on massless charged particles \cite{WW}. Of course, 
we now regard the 
masslessness of the gluons to be a {\em consequence\/} of gauge invariance, 
just as in QED.} 

\begin{figure}
\begin{center}
\epsfxsize=0.9\textwidth
\leavevmode
\epsfbox{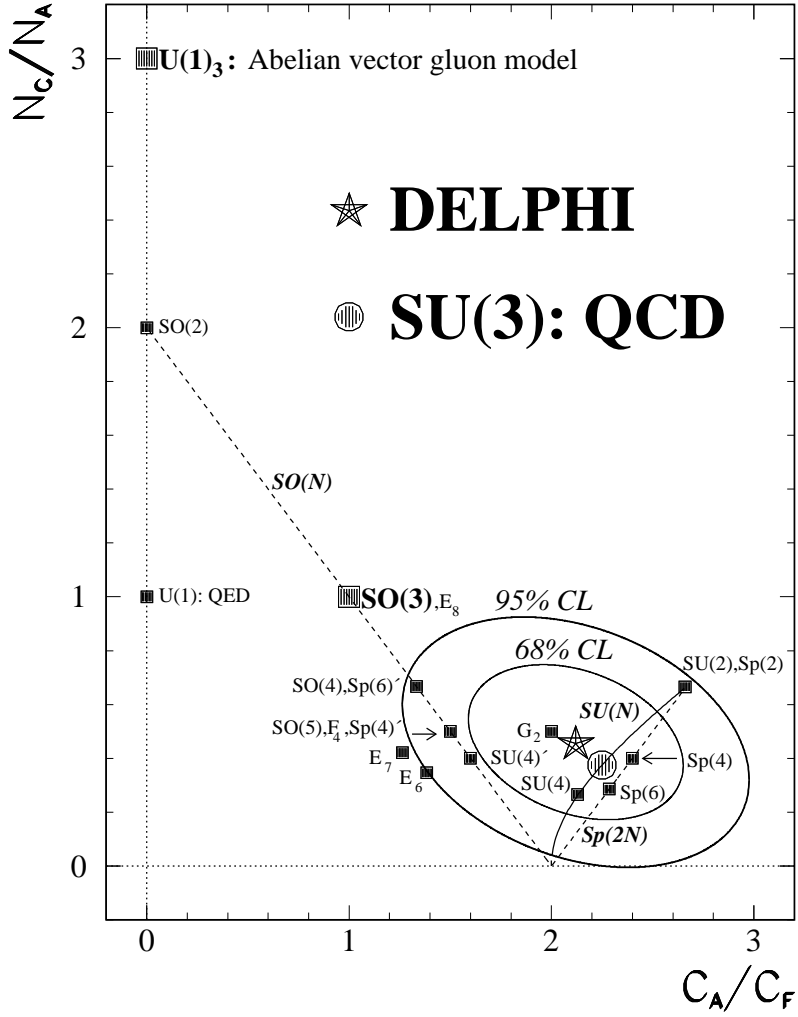}
\end{center}
\caption[fake]{Comparison of $Z\to 4j$ ($\star$) with the expectation of a 
Yang-Mills gauge theory based on various gauge groups ($\Box$) and SU(3) 
($\bigcirc$).  $N_C/N_A$ is the ratio of the number of quark colors to the 
number of gluons, and $C_A/C_F$ is the ratio of the strength of 
the three-gluon interaction to that of gluon bremsstrahlung from quarks.
Figure from Ref.~\cite{LEP}.}
\end{figure}

The gluon self interaction is believed to be responsible for much of the 
physics of QCD which sets it apart from QED, such as confinement
and asymptotic freedom. The gluon self interaction has been tested via $Z 
\to 4j$ events at LEP, where the decay $Z \to q\bar q gg$ involves the 
three-gluon interaction.  Since gauge invariance is mandatory for the 
consistency of the theory, it is not acceptable to arbitrarily vary the 
three-gluon interaction when comparing theory with experiment. Instead, 
the analysis by the LEP experiments leaves the Yang-Mills gauge symmetry 
intact, but varies the 
gauge group (leaving the fermions in the fundamental representation, 
as in QCD) \cite{LEP}.  Fig.~2 shows the result of an analysis of $Z\to 4j$, 
comparing the expectation of various gauge groups (boxes) and QCD (circle)
with the data (star); the axes identify the gauge group, and are 
explained in the figure caption.  The agreement of the data with the SU(3)
prediction is impressive.

As with QED, gauge symmetry does not mean there cannot be anomalous 
vector-boson self interactions.
The most general Lagrangian for gluons and quarks, consistent with 
Lorentz invariance and SU(3) gauge symmetry, is 
\begin{eqnarray}
{\cal L} & = & -\frac{1}{2}{\rm Tr}\,G^{\mu\nu}G_{\mu\nu} +i\bar\psi 
\not\!\! D\psi - m\bar\psi\psi \label{QCD} \\
         & + & \frac{c_1}{M^2}\bar\psi\gamma^\mu\psi\bar\psi\gamma_\mu\psi + 
         \frac{c_2}{M^2} {\rm Tr}\,G^\mu_\nu G^\nu_\rho G^\rho_\mu + \cdots  
	 \nonumber 
\end{eqnarray}
where $G^{\mu\nu}$ is the non-Abelian field-strength tensor.  The first 
line is the Lagrangian of ordinary QCD, and the (infinite number of) 
additional terms correspond to new physics associated with a mass scale 
$M$, as in QED.  The first such term corresponds to a four-quark contact 
interaction, and is searched for in high-$p_T$ jet events at the 
Tevatron, resulting in a lower bound on $M$ of about 1 TeV \cite{COMP}.  
The second such 
term yields an anomalous three-gluon interaction,\footnote{As mentioned 
in a previous footnote, the analogous term vanishes in QED. It does not 
vanish in QCD because $G^{\mu\nu}$ is an SU(3) matrix.} and is best sought in 
top-quark production at the LHC \cite{SIMMONS}.\footnote{This term may also
be sought in $Z\to 4j$, which yields a weak bound \cite{DZ}.}  It also
yields an anomalous four-, five-, and six-gluon interaction.

As with QED, there is no reason not to expect these additional terms in 
the Lagrangian to be
present, but there is also nothing which tells us at what mass scale, $M$,
we should expect them to manifest themselves. The renormalizability of 
ordinary QCD ensures that these terms are not necessary to cancel divergences,
to all orders in perturbation theory. However, as with QED, the 
renormalizability 
of the theory is simply a consequence of the fact that $M$ is much greater 
than the currently accessible energy and accuracy. 

\section*{Weak Interaction}

\indent\indent Let's move on to the weak interaction.  As with QED and 
QCD, we begin with experimental facts:
\begin{enumerate}

\item The weak bosons are massive.

\item The weak bosons have spin one.

\end{enumerate}
The big difference between the weak interaction and both QED and QCD 
is that the vector bosons are massive.  Let's 
construct a consistent theory of massive vector bosons, associated with a
field $W^\mu(x)$.  We'll add more experimental facts later.  

As with QED and QCD, we immediately run into the problem that the vector 
field $W^\mu$ has too many degrees of freedom.\footnote{At this point, I 
use 
$W^\mu$ to denote a generic massive vector field.}  This problem is less 
severe for a massive spin-one particle because it has three degrees of freedom,
corresponding to helicities $\pm 1,0$, rather than the two degrees of 
freedom of the massless case.  As with QED and QCD, the temporal
component of the vector field corresponds to a state of negative energy, so 
we must eliminate it as a dynamical degree of freedom.

Consider the following Lagrangian for a non-interacting vector field $W^\mu$
of mass $M_W$:
\begin{equation}
{\cal L} = -\frac{1}{4} W^{\mu\nu}W_{\mu\nu} + \frac{1}{2}M_W^2 W^\mu W_\mu 
\label{WEAK}
\end{equation}
where 
\begin{equation}
W^{\mu\nu}= \partial^\mu W^\nu - \partial^\nu W^\mu \;.
\end{equation}
The kinetic part of the Lagrangian is written in terms of the field-strength 
tensor $W^{\mu\nu}$ in order to remove $W^0$ as a dynamical field, just 
as we did for QED and QCD.  However, for the case of a massive vector 
field one can do even better; the extra degree of freedom can be removed 
in a manifestly Lorentz-invariant manner \cite{ROMAN,BS,IZ,R}. 
Consider the equation of motion of the 
field $W^\mu$, derived from the Lagrangian in Eq.~\ref{WEAK}:
\begin{equation}
\partial_\nu W^{\nu\mu} + M_W^2 W^\mu = 0 \;.
\label{EOM}
\end{equation}
Now apply $\partial_\mu$ to this equation.
The first term vanishes since $W^{\mu\nu}$ is antisymmetric, so we find
\begin{equation}
\partial_\nu W^\nu = 0 \;.
\label{CONSTRAINT}
\end{equation}
This is a constraint equation on the field $W^\mu$, and allows us to
remove one degree of freedom.  Since it is a Lorentz-invariant 
condition, Lorentz invariance remains manifest.\footnote{Alternatively, 
one can simply impose this constraint on the field as an auxiliary 
condition \cite{BS}.}

Although Eq.~\ref{CONSTRAINT} is reminiscent of the familiar Lorentz 
gauge condition of QED and QCD, it is not a gauge condition at all.  The 
massive vector theory has no gauge invariance whatsoever; gauge 
invariance is {\em non-existent\/} and {\em unnecessary}.  This is in striking 
contrast to QED and QCD. Because we are so used to working with gauge theories, 
this simple point 
is sometimes forgotten.  The quantization of a massless vector theory 
such as QED or QCD is a difficult task, and one tends to forget how easy 
it is to quantize a massive vector theory.\footnote{If the massive vector 
theory is a spontaneously-broken gauge theory, quantization is as 
complicated as in QED and QCD, of course.}

This construction is not upset by the introduction of interactions with 
other fields, or even self interactions. One way to see this is to 
consider the propagator of the vector field.  The free field equation, 
Eq.~\ref{EOM}, written in terms of the vector field, is
\begin{equation}
\Box W^\mu - \partial_\nu\partial^\mu W^\nu + M_W^2 W^\mu = 0 \;.
\end{equation}
This yields the momentum-space propagator
\begin{equation}
D^{\mu\nu}(p) = i\frac{-g^{\mu\nu}+\frac{p^\mu p^\nu}{M_W^2}}{p^2-M_W^2}\;.
\end{equation}
The numerator of the propagator contains the sum over the three 
polarization states corresponding to the three helicity states of a 
massive spin-one particle, and nothing more.\footnote{This is the familiar
``unitary gauge'' propagator of a spontaneously-broken gauge theory,
meaning it contains only the physical 
polarization states.  Since we are not (yet) treating the massive vector field 
as a gauge field, we avoid this language.}  Hence there is no concern 
about interactions potentially coupling to unphysical polarization states,
as there is in QED and QCD \cite{VELTMAN,BS,IZ}.\footnote{In QED and 
QCD, the propagator must couple to a conserved current, or gauge 
invariance (and hence Lorentz invariance) is lost \cite{BD}.}

Given that gauge invariance has nothing to do with a generic massive 
vector boson theory, one must wonder why we believe the weak interaction 
is described by a gauge theory.  The answer lies in a third experimental 
fact:
\begin{enumerate}
\setcounter{enumi}{2}
\item The couplings of the weak bosons to the three generations of quarks 
and leptons are, to high precision, those of an 
SU(2)$_L\times$U(1)$_Y$ gauge theory.

\end{enumerate}
But if the weak interaction is a gauge theory, why aren't the weak bosons 
massless, as appears to be required of gauge bosons?  The well-known solution 
to this puzzle is that the gauge symmetry is spontaneously broken \cite{WS}.  
This means that while the Lagrangian {\em is\/} invariant under 
SU(2)$_L\times$U(1)$_Y$
gauge transformations, the solution to the Lagrangian is not.

A skeptic might ask if the (local) gauge symmetry is really necessary.  
Wouldn't it be enough to impose {\em global\/} SU(2)$_L\times$U(1)$_Y$ symmetry 
on the Lagrangian to reproduce the observed couplings of the weak bosons 
to fermions?  The answer is no; one needs the local gauge symmetry to 
explain the universality of the weak interaction, i.e., to explain why 
the weak bosons couple the same to quarks as to leptons, and to all 
three generations (as far as we know).\footnote{It is particularly 
important that we 
test the universality of the weak interaction with respect to the 
recently-discovered top quark \cite{TOP}.}  To see this, consider the
Lagrangian for the coupling of the weak bosons to fermions,\footnote{Here 
and throughout I consider only the SU(2)$_L$ part of the weak 
interaction, and ignore the hypercharge interaction.  The field $W^\mu$ 
represents an SU(2)$_L$ triplet of gauge fields, and $\tau$ are the usual Pauli
matrices. As usual, $\psi_L \equiv \frac{1}{2}(1-\gamma_5)\psi$ denotes the 
left-chiral fermion field.}
\begin{equation}
{\cal L} = i\bar\psi_L\gamma^\mu(\partial_\mu+ig\frac{1}{2} \tau \cdot 
W_\mu)\psi_L\;.
\label{UNIV}
\end{equation}
Each term is separately 
invariant under global SU(2)$_L$ transformations, regardless of the value of 
$g$. However, both terms are needed to ensure invariance under {\em local\/} 
SU(2)$_L$ transformations $U(x)$, 
\begin{eqnarray}
\psi_L & \to & U\psi_L \\
\tau \cdot W^\mu & \to & U \tau \cdot W^\mu 
U^{\dagger} + \frac{2i}{g} (\partial^\mu U)U^{\dagger}
\end{eqnarray}
and they must be present exactly as shown in Eq.~\ref{UNIV}, with the same 
coupling $g$ for all fermions \cite{HAGI}.

The skeptic might counter that, while willing to accept
local gauge invariance as the explanation of the universality of the 
coupling of weak bosons to fermions, this 
universality need not extend to the gauge-boson self interactions.
Couldn't one imagine that the gauge symmetry is present only in the fermionic
sector of the theory?  The answer to this is also negative.  The one-loop 
correction to the coupling of weak bosons to fermions involves the 
weak-boson self interaction, and unless this interaction is of the 
Yang-Mills form, it will generally destroy the gauge-theory form of the 
fermionic coupling.  While the couplings may be ``readjusted'' to their 
experimentally-observed values, the explanation of universality is lost.  
This problem is especially severe in light of 
the fact that the quarks experience the strong interaction, while the 
leptons do not, so the amount of ``readjustment'' necessary will generally 
differ for the two types of fermions.  Thus we conclude that in order for
gauge symmetry to explain the universality of the weak interaction, it 
must be a symmetry of the full Lagrangian, not just part of it.

Just as in QED and QCD, anomalous vector-boson self interactions may be 
introduced via higher-dimension terms in the Lagangian, suppressed by 
inverse powers of some mass scale, $M$.  However, in the weak interaction, 
the implementation of this differs depending on whether or not a 
fundamental Higgs field is introduced in the Lagrangian.  Below we pursue 
these possibilities separately.

\subsection*{Higgs model}

\indent\indent Consider including the Higgs-doublet field, $\phi$, to break the 
electroweak symmetry in the standard way.  The Lagrangian is
\begin{eqnarray}
{\cal L} & = & -\frac{1}{8}{\rm Tr}\,W^{\mu\nu}W_{\mu\nu} +i\bar\psi_L \not 
	 \!\!D\psi_L \label{HIGGS} \\
         & + & (D^\mu\phi)^{\dagger}D_\mu\phi - V(\phi^{\dagger}\phi) 
	 \nonumber \\
         & + & \frac{c_1}{M^2}(D^\mu\phi)^{\dagger} W_{\mu\nu} D^\nu\phi +
         \frac{c_2}{M^2} {\rm Tr}\,W^\mu_\nu W^\nu_\rho W^\rho_\mu + \cdots  
	 \nonumber
\end{eqnarray}
where $W^{\mu\nu}$ is the full non-Abelian field-strength tensor,
\begin{equation}
W^{\mu\nu} = \tau \cdot (\partial^\mu W^\nu - \partial^\nu W^\mu 
- g W^\mu \times W^\nu)\;.
\end{equation}
The first two lines are the standard electroweak Lagrangian. 
When the Higgs field $\phi$ acquires a vacuum-expectation value, the 
first term in the third line produces additional three- and four-$W$ 
interactions.  The last term contributes additional three-, four-, 
{\mbox five-,} 
and six-$W$ interactions \cite{HAGI,HERN,W}.  These anomalous vector-boson 
self interactions are 
suppressed by inverse powers of some mass scale, $M$, which is the scale 
at which the ordinary electroweak theory ceases to be a valid description 
of nature.  As with QED and QCD, we have no reason not to expect that such 
terms are there. Since the standard electroweak theory is renormalizable,
these terms are not necessary to cancel divergences, to all orders in 
perturbation theory, so $M$ can be arbitrarily large.
However, radiative corrections to the Higgs vacuum-expectation
value diverge quadratically,\footnote{An equivalent argument is usually 
presented in terms of the Higgs mass.} so the value $v\approx 250$ GeV 
is natural only if there is new physics which cuts off the divergence at 
or below 1 TeV \cite{CHIV}. Thus naturalness of the Higgs model suggests 
that $M$ should not be greater than about 1 TeV.\footnote{One possibility
for new physics which cuts off the quadratic divergence is supersymmetry.}

\subsection*{No-Higgs model}

\indent\indent Although we believe that the electroweak interaction is a 
spontaneously-broken gauge theory, we do not know if the spontaneous 
symmetry breaking is the result of the vacuum-expectation value of a 
fundamental Higgs field.  If we insist that the theory be renormalizable 
and perturbative (weak coupling), then the only option is indeed 
the standard Higgs model \cite{CLT} and generalizations thereof, such as a 
two-Higgs-doublet model as employed in the supersymmetric standard model 
\cite{LAHANAS}.
However, we have no guarantee that nature is so kind as to provide us 
with a symmetry-breaking mechanism that can be analyzed perturbatively.

Whatever the symmetry-breaking mechanism, it must provide the three 
Goldstone bosons which are absorbed by the $W^{\pm}$ and $Z$ bosons to 
become massive.  A generic approach to the symmetry-breaking physics is 
then to introduce only these three Goldstone bosons into the Lagrangian, 
but no other fields \cite{HAGI,HERN,W,AB,V}.  Although the resulting theory is 
non-renormalizable, it should be a valid effective field theory at 
energies below the mass scale of the symmetry-breaking physics 
responsible for the Goldstone bosons.

Let us introduce the three Goldstone-boson fields $\pi$ via the
field\footnote{The choice of the symbol $\pi$ to denote the Goldstone bosons
is by analogy with the physical pion field, which is an (approximate) Goldstone
boson of spontaneously-broken chiral symmetry in QCD.}
\begin{equation}
\Sigma \equiv \exp[i \tau \cdot  \pi/v]
\end{equation}
where $v=2M_W/g$.  The Lagrangian is 
\begin{eqnarray}
{\cal L} & = & -\frac{1}{8}{\rm Tr}\,W^{\mu\nu}W_{\mu\nu} +i\bar\psi_L 
	 \not \!\!D\psi_L \label{NOHIGGS} \\
         & + & \frac{v^2}{4} {\rm Tr}\,(D^\mu\Sigma)^{\dagger}D_\mu\Sigma 
	 \nonumber \\
         & + & c_1 \frac{v^2}{M^2}({\rm Tr}\,(D^\mu\Sigma)^{\dagger}D_\mu
	\Sigma)^2
	 + c_2 \frac{v^2}{M^2}{\rm Tr}\,W^{\mu\nu}(D_\mu\Sigma)^{\dagger}
	D_\nu\Sigma + \cdots \nonumber
\end{eqnarray}
where
\begin{equation}
D^\mu\Sigma = (\partial^\mu + i\frac{g}{2} \tau \cdot W^\mu)\Sigma
\end{equation}
is the gauge-covariant derivative. The $\Sigma$ field transforms under 
SU(2)$_L$ as
\begin{equation}
\Sigma \to U\Sigma\;.
\end{equation}
The first line in Eq.~\ref{NOHIGGS} is the usual weak-interaction Lagrangian.
The second line is
is responsible for the $W$ mass, which is evident when it is expanded in 
terms of the $\pi$ fields:
\begin{equation}
\frac{v^2}{4}{\rm Tr}\,(D^\mu\Sigma)^{\dagger}D_\mu\Sigma = 
\frac{1}{2}\frac{g^2}{4}v^2 W^\mu W_\mu
+ \frac{1}{2}\partial^\mu \pi \cdot \partial_\mu \pi 
+ \frac{1}{2v^2} 
(\pi \cdot \partial^\mu \pi)(\pi \cdot \partial_\mu \pi) + \cdots
\label{MASS}
\end{equation}
The physical content of the theory is manifest in the unitary gauge, 
$\Sigma=1$, in which the Goldstone bosons are completely absorbed by the 
weak vector bosons, and disappear from the Lagrangian.  However, it is
convenient for our discussion (and for calculational purposes) to consider
a gauge in which the Goldstone bosons are present.  

\begin{figure}
\begin{center}
\epsfxsize=0.9\textwidth
\leavevmode
\epsfbox{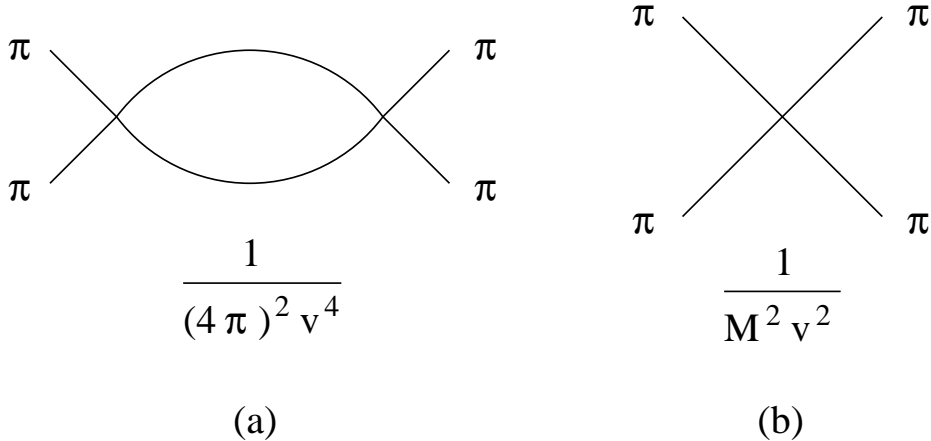}
\end{center}
\bigskip\bigskip
\caption{(a) One-loop amplitude for four Goldstone bosons.
(b) Tree-level interaction of four Goldstone bosons required to cancel
the divergence from the one-loop amplitude.}
\end{figure}

The first term in the third line contributes an anomalous four-$W$ 
interaction, and the second term an anomalous three- and four-$W$ interaction. 
These terms are suppressed by inverse powers of a mass scale, $M$, 
which is the scale at which the physics responsible for spontaneous 
symmetry breaking resides.  At first sight this mass scale can be made 
arbitrarily large, as in 
QED, QCD, and the Higgs model.  However, the term responsible for the 
vector-boson masses, Eq.~\ref{MASS}, contains a non-renormalizable 
four-$\pi$ interaction with a coefficient proportional to $1/v^2$, as shown.  
This coefficient sets the scale for the other non-renormalizable terms.  
A one-loop four-$\pi$ amplitude
constructed from two of these four-$\pi$ interactions, shown in Fig.~3(a), 
is of order 
$1/(4\pi)^2 v^4$, where the factor $1/(4\pi)^2$ arises from the loop 
integration. This has the same dimensions as the contribution to the 
four-$\pi$ amplitude from the terms in the last line of Eq.~\ref{NOHIGGS},
shown in Fig.~3(b),
of order $1/M^2v^2$.  Since the one-loop amplitude is divergent, these terms
must be there to cancel the divergence. Thus $M$ must be of order 
$4\pi v \approx$ 3 TeV or less.  
The physics responsible for electroweak symmetry breaking must therefore
manifest itself by at least 3 TeV. 

There is one last experimental fact we can add to our 
discussion:
\begin{enumerate}
\setcounter{enumi}{3}
\item $M_W \approx M_Z \cos\theta_W$

\end{enumerate}
This is embodied in the 
$\rho$ parameter,
\begin{equation} 
\rho\equiv M_W^2/(M_Z^2 \cos^2\theta_W) \approx 1\;.
\label{RHO}
\end{equation}
In other words, not 
only are the $W$ and $Z$ bosons massive, but their masses are related.  
This can be explained by hypothesizing that the symmetry-breaking sector 
of the theory possesses a global SU(2) symmetry, called a ``custodial'' 
symmetry \cite{CHIV,V,TECH,SSVZ}. Although it is not always made explicit, the 
standard Higgs model contains this symmetry.  Models of dynamical electroweak 
symmetry breaking, such as Technicolor \cite{CHIV,TECH}, must contain this 
symmetry even if Eq.~\ref{RHO} is satisfied at tree level, since the 
corrections are potentially large (strong coupling).  The custodial symmetry
is further evidence that the properties of the weak interaction are dictated
by symmetry.

\section*{Outlook}

Of the electromagnetic, strong, and weak interactions, only the last 
guarantees new physics at an accessible mass scale.  This is the physics 
associated with electroweak symmetry breaking. All we know for sure about 
this physics is that it must manifest itself by at least $4\pi v\approx 3$ TeV.
Furthermore, the fact that 
the $W$ and $Z$ masses are related by $M_W \approx M_Z \cos\theta_W$ suggests 
that the symmetry-breaking sector contains a ``custodial'' global SU(2) 
symmetry.

The electroweak-symmetry-breaking sector could be the 
source of anomalous weak-vector-boson self interactions. 
We have considered two scenarios for the electroweak-symmetry-breaking 
physics:
\begin{enumerate}

\item Higgs model

\item No-Higgs model: only Goldstone bosons up to $4\pi v \sim 3$ TeV

\end{enumerate}
These two models are so commonly studied that one begins to think they 
are the only possibilities.  This is not the case. The symmetry-breaking
physics could be very rich, containing resonances, new fermions, new 
gauge bosons, etc.  The fact that nature makes use of gauge theories for 
the three known low-energy forces leads one to guess that the 
symmetry-breaking sector is also a gauge theory.  Examples which 
implement this idea are fixed-point Technicolor \cite{H}, walking
Technicolor \cite{WALK}, two-scale
Technicolor \cite{EL}, fermions in large representations of the gauge group
\cite{M,SW}, etc.  From a 
theoretical point of view, these models receive less attention because 
they are neither amenable to a perturbative analysis (like the Higgs model)
nor to a ``model-independent'' analysis (like the No-Higgs model). 
They also 
generically run into difficulty with precision electroweak experiments 
\cite{SCHAILE,T,PT}, something we have learned from the vector-boson era. 
Nature may not care about any of these objections.  We should probe 
higher energies and keep an open mind regarding the 
manner in which the physics of electroweak symmetry breaking reveals 
itself.

\begin{figure}
\begin{center}
\epsfxsize=0.9\textwidth
\leavevmode
\epsfbox{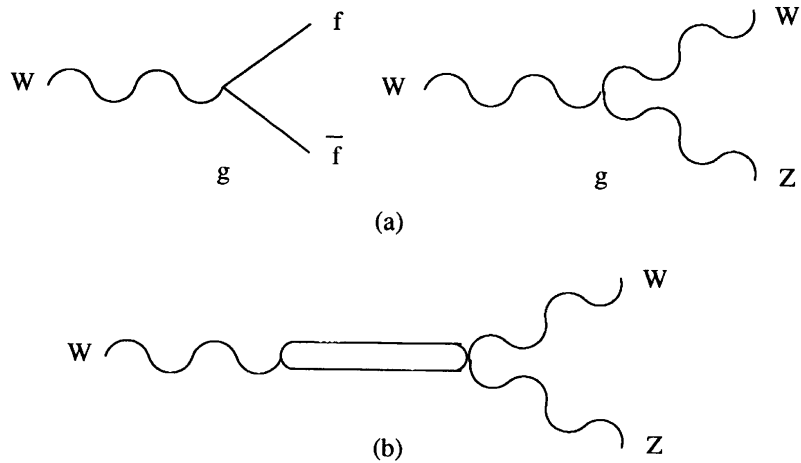}
\end{center}
\bigskip\bigskip
\caption{(a) Schematic illustration of the universality of the 
coupling of weak bosons to fermions and to themselves.
(b) An anomalous weak-boson self interaction produced by a
$J=1$ resonance.}
\end{figure}

After this long discussion, we are now prepared to go back and answer a
question we posed at the beginning: What can we learn from the era of weak 
boson pair production?  At the very least, we will see a confirmation of 
the universality of the weak interaction, extended to the weak-boson self 
interaction, as depicted schematically in Fig.~4(a). It is the universality 
of the fermionic couplings of the weak bosons which led us to 
the electroweak theory in the first place, so this confirmation will be a
crowning achievement.  However, we hope for much more from this era; we 
anticipate at least the first signs of the physics responsible for electroweak 
symmetry breaking, and, at best, the complete revelation of this physics.
One possible manifestation of this physics is a $J=1$ resonance which couples 
to the weak bosons, as depicted in Fig.~4(b).  Although we would not 
usually regard this as an ``anomalous vector-boson self interaction'', 
there is no reason why we should not.  If we observe such a 
resonance, it would be a {\it very} anomalous vector-boson self 
interaction.

\section*{Acknowledgments}

I am grateful for conversations and correspondence with T.~Appelquist,
W.~Bardeen, C.~Burgess,
D.~Dicus, E.~Eichten, A.~El-Khadra, W.~Marciano, D.~Morris, C.~Quigg, 
J.~Stack, G.~Valencia, S.~Weinberg, J.~Wudka,
D.~Zeppenfeld, and D.~Zwanziger.  This work was supported in part
by Department of Energy grant DE-FG02-91ER40677.


\begin{thebibliography}{99}

\bibitem{WZ} UA1 Collaboration, G.~Arnison~{\it et al.}, \PLB 122B 103 1983 ;
{\bf 126B}, 398 (1983); {\bf 129B}, 273 (1983); UA2 Collaboration, 
M.~Banner~{\it et al.}, \PLB 122B 476 1983 ; P.~Bagnaia~{\it et al.}, \PLB 129B 
130 1983 .

\bibitem{NOBEL} S.~van~der~Meer, \RMP 57 689 1985 ; 
C.~Rubbia, \RMP 57 699 1985 .

\bibitem{FUESS} T.~Fuess, these proceedings.

\bibitem{CDF} CDF Collaboration, F.~Abe~{\it et al.}, 
CDF/ANAL/ELECTROWEAK/CDFR/\\2951 (1995).

\bibitem{BUS} J.~Busenitz, these proceedings.

\bibitem{WOM} J.~Womersley, these proceedings.

\bibitem{BAR} T.~Barklow, these proceedings.

\bibitem{PHYSICA} S.~Weinberg, Physica~{\bf 96A}, 327 (1979).

\bibitem{EFT} The concept of ``effective'' quantum field theory has become 
widespread, and there are many excellent introductions.  Among them are
G.~P.~Lepage, in {\it From Actions to Answers, Proceedings of the 1989 
Theoretical Advanced Study Institute (TASI)}, eds. T.~DeGrand and 
D.~Toussaint (World Scientific, Singapore, 1990), p.~483; 
S.~Weinberg, in {\it Proceedings of the XXVI
International Conference on High-Energy Physics}, ed. J.~Sanford 
(American Institute of Physics, New York, 1993), p.~346;
J.~Polchinski, in {\it Recent Directions in Particle Theory, Proceedings of 
the 1992 Theoretical Advanced Study Institute (TASI)}, eds. J.~Harvey and 
J.~Polchinski (World Scientific, Singapore, 1993), p.~235;
H.~Georgi, Ann.~Rev.~Nucl.~Part.~Sci.~{\bf 43}, 209 (1993).

\bibitem{EIN} M.~Einhorn, in {\it Workshop on Physics and Experiments with 
Linear $e^+e^-$ Colliders}, Hawaii, 1993, eds. F.~Harris, S.~Olsen, S.~Pakvasa,
and X.~Tata (World Scientific, Singapore, 1993), p.~122.

\bibitem{W5} S.~Weinberg, \PR 138 B988 1965 ; in {\it 1964 Brandeis Summer
Institute in Theoretical Physics, Lectures on Particles and Field Theory}
(Prentice-Hall, Englewood Cliffs, 1965), Vol.~2, p.~405.

\bibitem{BD} J.~Bjorken and S.~Drell, {\it Relativistic Quantum Fields}
(McGraw-Hill, New York, 1965).

\bibitem{Z} D.~Zwanziger, \PR 133 B1036 1964 . 

\bibitem{VELTMAN} An engaging and non-technical discussion is given in 
 M.~Veltman, {\it Diagrammatica} (Cambridge University Press, Cambridge, 1994).

\bibitem{3JET} DELPHI Collaboration, P.~Abreu~{\it et al.}, \PLB 274 498 1992 ;
L3 Collaboration, B.~Adeva~{\it et al.}, \PLB B263 551 1991 ;
OPAL Collaboration, G.~Alexander~{\it et al.}, \ZPC 52 543 1991 .

\bibitem{WW} S.~Weinberg and E.~Witten, \PLB 96B 59 1980 .

\bibitem{COMP} CDF Collaboration, F.~Abe~{\it et al.}, \PRL 71 2542 1993 ;
{\bf 74}, 3538 (1995).

\bibitem{SIMMONS} E.~Simmons, these proceedings.

\bibitem{DZ} A.~Duff and D.~Zeppenfeld, \ZPC 53 529 1992 .

\bibitem{LEP} ALEPH Collaboration, D.~Decamp~{\it et al.}, \PLB B284 151 1992 ;
DELPHI Collaboration, P.~Abreu~{\it et al.}, \ZPC 59 357 1993 ;
L3 Collaboration, B.~Adeva~{\it et al.}, \PLB B248 227 1990 ;
OPAL Collaboration, R.~Akers~{\it et al.}, \ZPC 65 367 1995 .

\bibitem{ROMAN} P.~Roman, {\it Introduction to Quantum Field Theory} 
(Wiley, New York, 1969).

\bibitem{BS} N.~Bogoliubov and D.~Shirkov, {\it Introduction to the 
Theory of Quantized Fields} (Interscience, New York, 1959).

\bibitem{IZ} C.~Itzykson and J.-B.~Zuber, {\it Quantum Field Theory}
(McGraw-Hill, New York, 1980).

\bibitem{R} L.~Ryder, {\it Quantum Field Theory} (Cambridge University Press,
Cambridge, 1985).

\bibitem{WS} S.~Weinberg, \PRL 19 1264 1967 ; A.~Salam, in {\it Elementary
Particle Theory: Relativistic Groups and Analyticity (Nobel Symposium No.~8)},
ed. N.~Svartholm (Almqvist and Wiksell, Stockhom, 1968), p.~367.

\bibitem{TOP} CDF Collaboration, F.~Abe~{\it et al.}, \PRL 74 2626 1995 ;
D0 Collaboration, S.~Abachi~{\it et al.}, \PRL 74 2632 1995 .

\bibitem{HAGI} K.~Hagiwara, these proceedings.

\bibitem{HERN} P.~Hernandez, these proceedings.

\bibitem{W} J.~Wudka, these proceedings.

\bibitem{CHIV} S.~Chivukula, these proceedings.

\bibitem{CLT} J.~M.~Cornwall, D.~Levin, and G.~Tiktopoulos, \PRL 30 1268 1973 ;
\PRD 10 1145 1974 ; C.~Llewellyn-Smith, \PLB B46 233 1973 .

\bibitem{LAHANAS} A.~Lahanas, these proceedings.

\bibitem{AB} T.~Appelquist and C.~Bernard, \PRD 22 200 1980 .

\bibitem{V} G.~Valencia, these proceedings.

\bibitem{TECH} S.~Weinberg, \PRD 19 1277 1979 ; 
L.~Susskind, \PRD 20 2619 1979 . 

\bibitem{SSVZ} P.~Sikivie, L.~Susskind, M.~Voloshin, and V.~Zakharov, 
\NPB B173 189 1980 .

\bibitem{H} B.~Holdom, \PRD 24 1441 1981 .

\bibitem{WALK} B.~Holdom, \PLB B150 301 1985 ;
K.~Yamawaki, M.~Bando, and K.~Matumoto, \PRL 56 1335 1986 ;
T.~Appelquist, D.~Karabali, and L.~Wijewardhana, \PRL 57 957 1986 ;
T.~Appelquist and L.~Wijewardhana, \PRD 36 568 1987 .

\bibitem{EL} E.~Eichten and K.~Lane, \PLB B222 274 1989 .

\bibitem{M} W.~Marciano, \PRD 21 2425 1980 .

\bibitem{SW} S.~Willenbrock, \PLB B340 236 1994 .

\bibitem{SCHAILE} D.~Schaile, these proceedings.

\bibitem{T} T.~Takeuchi, these proceedings.

\bibitem{PT} M.~Peskin and T.~Takeuchi, \PRL 65 964 1990 ;
B.~Holdom and J.~Terning, \PLB B247 88 1990 ;
M.~Golden and L.~Randall, \NPB B361 3 1991 .

\end{thebibliography}
\end{document}